\documentclass[twocolumn,prl,aps,amsmath,amssymb,color,superscriptaddress]{revtex4}
\usepackage{stmaryrd}
\usepackage{amsmath}
\usepackage{amssymb}
\usepackage{graphicx}
\usepackage{dcolumn}
\usepackage{bm}
\usepackage{tabularx}
\usepackage{color}
\begin{document}
\begin{titlepage}
\title{Polarization at the Nanoscale}
\author{Ke Yang}
\affiliation{Department of Physics, Applied Physics and Astronomy, Rensselaer Polytechnic Institute, Troy, New York 12180 (USA)}
\author{Zeyu Jiang}
\affiliation{Department of Physics, Applied Physics and Astronomy, Rensselaer Polytechnic Institute, Troy, New York 12180 (USA)}
\author{Duk-Hyun Choe}
\affiliation{Samsung Advanced Institute of Technology (SAIT), Suwon, Korea}
\author{Damien West}
\email{damienwest@gmail.com}
\affiliation{Department of Physics, Applied Physics and Astronomy, Rensselaer Polytechnic Institute, Troy, New York 12180 (USA)}
\author{Shengbai Zhang}
\affiliation{Department of Physics, Applied Physics and Astronomy, Rensselaer Polytechnic Institute, Troy, New York 12180 (USA)}
\date{\today}

\begin{abstract}
Modern polarization theory yields surface bound charge associated with spontaneous polarization of bulk. However, understanding polarization in nano systems also requires a proper treatment of charge transfer between surface dangling bonds. Here, we develop a real-space approach for total polarization and apply it to wurtzite semiconductors and $BaTiO_3$ perovskite. First-principles calculations utilizing this approach not only yield spontaneous bulk polarization in agreement with Berry phase calculations, but also uncover phenomena specific to nano systems. As an example, we show surface passivation leads to a complete quenching of the piezoelectric effect, which remerges only at larger length scale and/or spontaneous polarization.
\end{abstract}

\maketitle
\draft
\vspace{2mm}
\end{titlepage}
Macroscopic polarization of bulk materials is a fundamental concept in condensed matter physics. Not only does it serve as the origin for a number of fascinating physical phenomena such as the integer and fractional quantum Hall effects\cite{1,2}, superconductivity at heterostructures\cite{3}, but it is also the source of internal polarization fields\cite{3,4}, resulting in band bending\cite{5}. Such internal fields can boost photo-excited functionalities by enhancing photo-carrier separation in photo-catalysis\cite{6,7}, photovoltaics\cite{8}, and other optoelectronic applications\cite{9}. In addition, polarization plays an important role in surface doping\cite{10}, polar surface treatment\cite{4}, and polar surface stability\cite{3,11}. Understanding of the polarization in solids is of both fundamental and practical importance.

Bulk polarization can be understood using the theory of King-Smith and Vanderbilt\cite{12,13}, where the spontaneous polarization of a bulk material is cast in terms of a Berry-phase integral in reciprocal space. 
Despite the lack of explicit surfaces, in the bulk limit the surface bound charge ($\sigma_b$) can be deduced\cite{12} which in turn yields the surface free charge ($\sigma_f$), such as those arising from surface dangling bonds (DBs), as they must compensate ($\sigma_b$+$\sigma_f=0$) due to the vanishing bulk internal field. Note that if even a tiny field were present in bulk, the potential offset across a macroscopic solid would be gigantic, much larger than any band gap.
 In the case of nanostructures, however, there is generally not a complete cancellation between the bound and free charges, leading to a total polarization that could be larger than the spontaneous polarization. While the bound charge can be inferred from the Berry phase calculations, there is currently no rigorous method of calculating surface free charge for finite systems. This is because at the microscopic level, it has not been possible to uniquely separate the surface charge from bulk charge, even in non-polar systems\cite{15}. This is especially problematic for nanostructures due to the increased surface to bulk ratio.

At nanoscale, spontaneous polarization, surface chemistry, and size, all have profound effects on the internal field associated with total polarization. For the case of (fully) passivated surfaces, at small length scales the spontaneous polarization leads to a constant internal E-field, given simply by $P_{SP}/\epsilon_0$ where $P_{SP}$ is the spontaneous polarization of bulk. However, for $L>E_g\epsilon_0/P_{SP}$ such a constant field would lead to the valence band maxima (VBM) of one side of the material to be above the conduction band minima (CBM) of the opposing side, resulting in a charge transfer and accumulation of free charge, $\sigma_{\text {free }}$, as shown in Fig. 1(a). The situation can be more involved though, when DBs are present on surfaces. When the density of surface DBs is high, not all the electrons in donor DBs will be transferred to acceptor DBs on the opposing side of the material. Instead, there is a Fermi level line up between them, as schematically shown in Fig. 1(b). In physical systems, most often partial or nearly complete passivation of surface DBs take place. For instance, in a case where it is energetically favorable for the surface DB’s to be passivated, but entropy ensures some low density of active sites, as depicted in Fig. 1(c), a nearly complete charge transfer between DBs could be favorable, as the resulting field would be small. Here, there would be some initial free surface charge density, $\sigma_{\text {free }}^{0}$, which depends on the active defect density and remains constant until the potential across the sample brings the DB states into degeneracy, i.e. $\frac{\left(\sigma_{free}^{0}+\sigma_{SP}\right) L}{\epsilon_{0}}=E_{g}^{db}$, where $E_g^{\text{db}}$ is the energy difference between the DB states in zero field.

On the experimental front, size-dependent permanent dipole moment and those induced by surface strain or surface charges have been observed in wurtzite nanostructures\cite{6,16,17}. Unlike its zincblende counterpart, the wurtzite structure has intrinsic spontaneous polarization due to its low crystal symmetry. Size-dependent polarization field of $PbTiO_3$ nanomaterials has also been demonstrated and been used to improve photocatalytic activity\cite{7}. At the interface between $KNbO_3$ and $BaTiO_3$ and between $LaMoO_3$ and $SrMnO_3$, there supposed to exist the so-called polar catastrophe due to spontaneous polarization, which is, however, quenched\cite{18,19}. The quench may be explained as a band-bending effect: namely, bulk CBM lines up with VBM, when the film thickness $L$ exceeds a critical thickness $L_c$. Or it happens because surface DBs come into play. Most recently, permanent dipoles were also observed in zincblende $CdSe$ nano-platelets by transient electric birefringence measurement\cite{20}. Importantly, zincblende is a high-symmetry structure which should not have any spontaneous polarization due to bound charge. Clearly, none of the above can be explained in terms of a spontaneous polarization of bulk. An accurate microscopic theory for finite and nano systems is thus timely and highly desirable. Ideally, the bulk charge density may be used as a reference in the calculation of polarization, as it satisfies the zero macroscopic-electric-field requirement\cite{21}. However, arbitrariness in the choice of unit cell for bulk systems have so far rendered such a reference useless. 

In this paper, we develop a theory of polarization charge, from which to derive polarization at the nanoscale. We show that not only can the polarization charge be uniquely defined, but also its spatial distribution along the normal direction to the surfaces can be obtained. The approach is applied to seven wurtzite (I-VII, II-VI, III-V, and IV-IV) semiconductor slabs, as well as the perovskite $BaTiO_3$. We see that for passivated surfaces the total polarization charge is consistent with what is expected from Berry-phase calculations. For unpassivated surfaces at the nanoscale, however, charge transfer between DBs can dominate the total polarization. Furthermore, even when surface DBs are fully passivated, the entire system is semiconducting only when the polarization is relatively weak, and the length scale is small so that charge transfer between valance and conduction bands at opposing ends of the materials does not take place. In this semiconducting regime, piezeoelectric effects are quenched in qualitative agreement with experiment.

\begin{figure}[tbp]
\includegraphics[width=0.9\columnwidth]{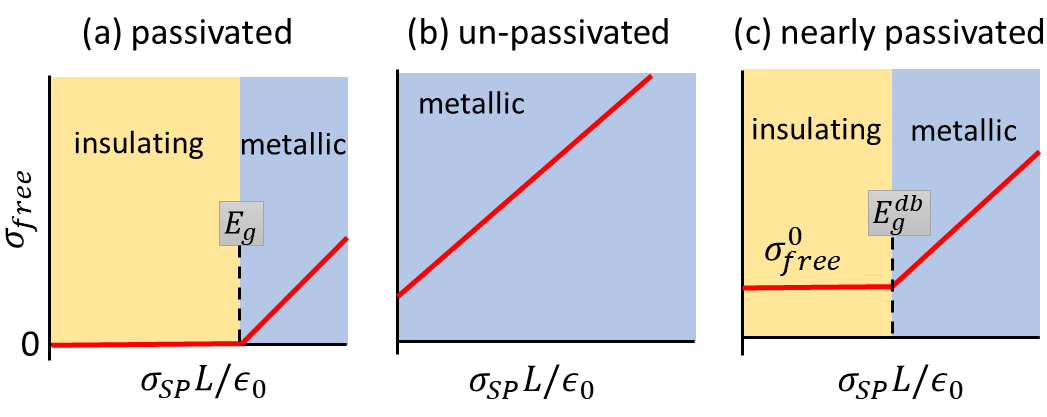}
\caption{\label{fig:fig1}  Schematic dependence of surface free charge density, $\sigma_{free}$, on spontaneous polarization, $\sigma_{SP}$, and sample size, $L$, for a polar material: (a) passivated surface, (b) unpassivated surface with a high concentration of DBs, and (c) nearly passivated surface with a low concentration of DBs.}
\end{figure}

To determine the spontaneous polarization in real space (and distinguish it from the free polarization), we need a meaningful reference system with no internal electric field. Here, we use the one from Ref.\cite{15}, which fulfills the zero internal field requirement. When dealing with a macroscopic electric field normal to the surface of a slab, only the one-dimensional charge distribution $\rho(z)$ [Fig. 2(a)] matters. For this reason, we define the planar-averaged polarization charge density
\begin{equation}
\delta\rho(\mathrm{z})=\rho(\mathrm{z})-\rho_{0}(\mathrm{z})
\end{equation}     
where $\delta\rho(\mathrm{z})$ is the charge density of the reference bulk [in Fig. 2(b)], which is truncated at the unit cell boundary and hence goes to zero abruptly. Figure 2(c) depicts schematically $\delta\rho(\mathrm{z})$, the black line, which has contributions from both (i) free surface charge associated with surface DBs (the red line) and (ii) bound surface charge due to (intrinsic) spontaneous polarization of bulk (the blue line). The corresponding electrostatic potentials of $\delta\rho(\mathrm{z})$, $\rho_{0}(\mathrm{z})$, and $\delta\rho(\mathrm{z})$ are shown in Figs. 2(d)-(f), respectively. The total polarization can be obtained through integration of $\delta\rho(\mathrm{z})$, which reveals the net transfer of charge from one side of the slab to the other.

To directly compare the polarization with that from a Berry-phase calculation, however, it should be noted that $\sigma$ in a nanostructure includes screening effects, namely,
\begin{equation}
\sigma_{net}=\vec{P}_{net}\cdot\vec{n}
\end{equation}  
where
\begin{equation}
\vec{P}_{net}=\vec{P}_{SP}+\vec{P}_{ind}=\vec{P}_{sp}+\epsilon_{0}\chi\vec{E}
\end{equation}
In Eqs. (2) and (3), $\vec{P}_{net}$ is the net polarization, which includes both spontaneous polarization ($\vec{P}_{net}$) and induced polarization ($\vec{P}_{ind}$) and $\vec{n}$ is a unit vector normal to the surface. Note also that in Eq. (3) $\vec{P}_{ind}=\epsilon_{0}\chi\vec{E}$ where $\epsilon_{0}$, $\chi$, and $\vec{E}$ are the dielectric constant of vacuum, polarizability of bulk, and induced electric field, respectively.
As noted in Fig. 1(a), for passivated surfaces with small $L$ the slab is an insulator with no free surface charge. Here we note that our reference system is that of truncated bulk, in which by construction the one-dimensional macroscopic fields, $\vec{E}$, $\vec{P}$,and $\vec{D}$ all vanish. As the charge relaxation process only involves bound charge relaxation, this leads to change in $\vec{E}$ and $\vec{P}$, but not $\vec{D}$, Therefore $\vec{D}=0$, even after charge relaxation, leading to
\begin{equation}
\vec{D}=\epsilon_0\vec{E}+\vec{P}_{net}=0
\end{equation}
In the slab geometry, all the vectors here are in the same direction, i.e., the surface normal. Hence, we can drop the vector notations from Eqs. (2)-(4). By solving these equations, we arrive at
\begin{equation}
\vec{P}_{SP}=(1+\chi)\vec{P}_{net}=\epsilon_r\vec{P}_{net}
\end{equation}
where $\epsilon_r$ is the relative dielectric constant. 
\begin{figure}[tbp]
\includegraphics[width=0.9\columnwidth]{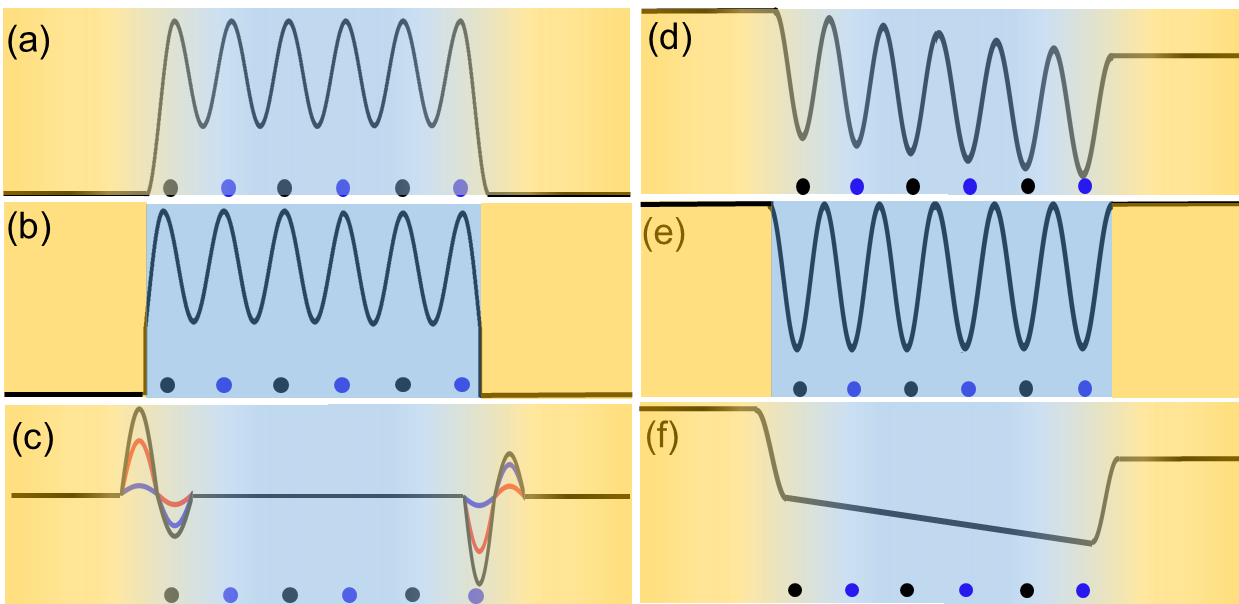}
\caption{\label{fig:fig2} Schematic diagrams of (a)-(c) planar-averaged charge densities and their corresponding electrostatic potentials (d)-(f), respectively. (a) $\rho(z)$ of a slab; (b) $\rho_0(z)$ of the corresponding truncated bulk; and (c) the total polarization charge $\delta\rho(z)=\rho(z)-\rho_{0}(z)$, shown in black, decomposed into the surface free charge and bound charge, shown in red and blue, respectively. The black and blue dots denote atomic positions of cations and anions.}
\end{figure}

To compare the results of this formulation of the real-space polarization and that of Berry phase, the polarization of a series of wurtzite semiconductor slabs were calculated self-consistently with density functional theory using the projector-augmented wave\cite{22} method, as implemented in the Vienna Ab Initio Simulation Package\cite{23,24}. The Perdew-Burke-Ernzerhof\cite{25} exchange-correlation functional was employed. The kinetic energy cutoﬀ was 500 eV and the Brillouin zone integration was performed on a $12 \times 12\times 12$ k-point grid (centered at $\Gamma$). To get an accurate real-space charge density, the total energy has been converged to within $10^{-8}$ eV and a dense grid in the FFT mesh along the polarization direction (with more than 200 points per \AA ) was used. For comparison, the bulk spontaneous polarization was calculated using the charge centers of Wannier functions\cite{26}, based on the Berry-phase approach\cite{13}. In the real-space approach, to isolate spontaneous polarization from the polarization due to surface free charge, surface dangling bonds are passivated according to the electron counting model (ECM)\cite{27,28} (passivation details for BN shown in Fig. S1 of the supplemental material).
\begin{table}
\centering
\caption{Calculated net polarization $\vec{P}_{net}$, dielectric constant $\epsilon_r$, spontaneous polarization $\epsilon_r\vec{P}_{net}$, [cf. Eq. (5)], and $\vec{P}_{SP}^{BF}$ calculated by the Berry-phase approach. Literature values are given in the last column. Except $\epsilon_r$, all quantities here are given in unit of $10^{-3} c/{m^2}$.}
\renewcommand\arraystretch{1.0}
\begin{ruledtabular}
\begin{tabular}{lcccccccccccccccccccccccccc}
       Substance & $\vec{P_{net}}$ & $\epsilon_r$ & $\epsilon_r\vec{P}_{net}$ &$\vec{P}_{SP}^{BF}$ & $\vec{P}_{SP}^{BF}(literature)$ \\
\hline
$SiC$	& -8.2 & 7.31	&-59 &	-45 & -15 \cite{31}; -40 \cite{10}
 \\
$BN$	& -4.3 &	4.47 &	-19 &	-18&   -12 \cite{32} \\
$AlN$	& -18	& 4.90	& -88 &	-87 &  -81 \cite{32}; -87 \cite{31}\\
$BeO$	& -12 &	3.21	& -38 &	-36 & -45 \cite{33}; -37 \cite{34}; -50 \cite{35} \\
$ZnO$	& -7.3 & 	5.80 & -42	& -37 & -57 \cite{33}; -5 \cite{35} \\
$CdSe$ &	-0.75	& 9.54	&  -7.4 &	-7.5 & -6 \cite{30}; -2 \cite{16}  \\
$CuCl$ &	-1.1 &	7.59 & -8.3 &	-6.2 &  - \\
\end{tabular}
\end{ruledtabular}
\end{table}

\begin{figure}[tbp]
\includegraphics[width=0.9\columnwidth]{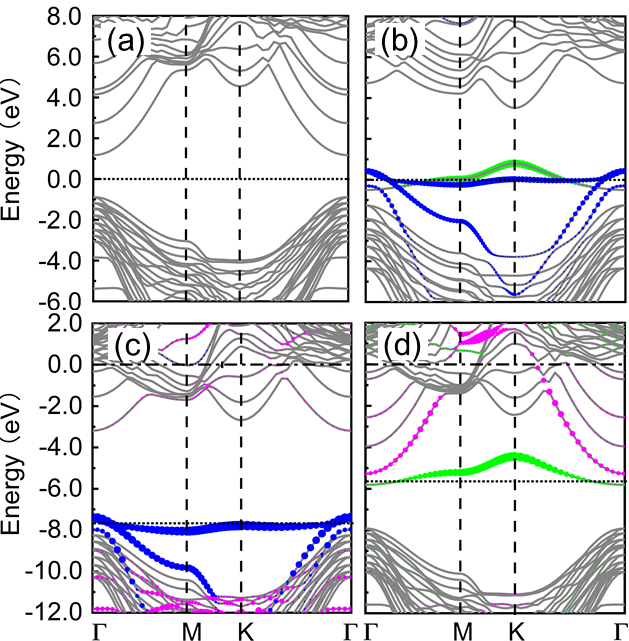}
\caption{\label{fig:fig3} Band structures of an 8-ML $BN$ slab: (a) Pseudo-H passivation; (b) no surface passivation. Here, black dotted line is the Fermi level and is set to zero. Blue, green bands are those from N and B surface states, respectively; (c) only the B-terminated surface is passivated; and (d) only the N-terminated surface is passivated. Magenta bands are pseudo-H states. In (c) and (d), vacuum levels outside the un-passivated surfaces are set to zero and are lined up.}
\end{figure}

Comparison of $P_{SP}$ calculated for seven different wurtzite semiconductors, ranging from I-VII, II-VI, III-V, to IV-IV, are provided in Table 1 and demonstrates that our non-Berry-phase real-space approach captures the essence of spontaneous polarization. In addition to providing a real-space framework to calculate the spontaneous polarization, this technique provides the ability to distinguish between bound and surface charge densities and the effects of surface DBs, as well as sample size, on the total polarization. For instance, in the case of un-passivated systems, the slab will become metallic and possess free surface charge, as electrons in the partially-occupied surface DBs on the opposing sides of the slab equilibrate with each other. Taking the wurtzite $BN(0001)$  slab in Fig. 3 as an example: with passivation, a gap opens; without passivation, electrons will transfer across the slab between the $B-BN(0001)$ and $N-BN(000\overline{1})$ surfaces until their quasi-Fermi levels line up with each other, as can be seen in Fig. 3(b). One may wonder what the quasi-Fermi level positions before the charge transfer would be. To see this, we perform calculations with one side, and only one side, of the slab passivated. This leaves the DB state of interest intact. Figures 3(c) and (d) show the resulting band structures for B- and N-terminated surfaces, respectively, in which we align the vacuum levels outside the un-passivated surfaces to offer an estimate of the quasi-Fermi level offset $\phi$. 

\begin{figure}[tbp]
\includegraphics[width=0.9\columnwidth]{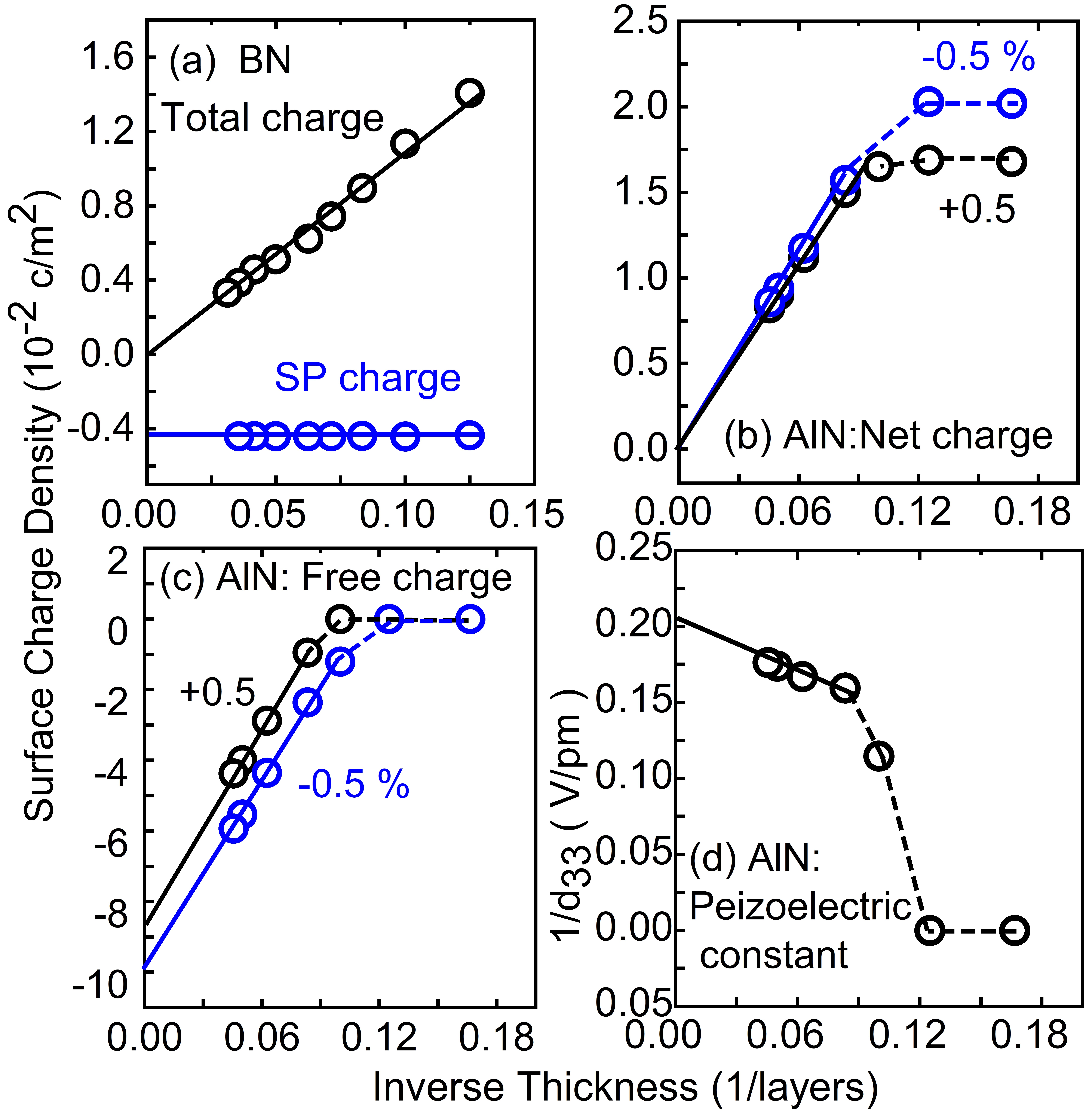}
\caption{\label{fig:fig4} (a)-(c) Charge density and (d) piezoelectric constant as functions of the inverse of slab thickness. (a) $BN$ and (b)-(d) $AlN$ where blue regions are metallic but yellow regions are insulating. In (b) and (c), black symbol and line are for a (compressive)  -0.5$\%$ strain, while red symbol and line are for a (tensile) 0.5$\%$ strain, along the normal $<0001>$ direction}
\end{figure}

For $BN$, both the DB states are made of atomic 2 $p$ orbitals. However, N is more electronegative than B and as such, the DB state of N should is lower in energy than that of B, as shown in Figs. 3(c) and (d), which yields a calculated $\phi$ of 2.4 eV for our 8 ML slab.
While an ionic model \cite{4} with ECM would predict a fixed charge transfer of $\frac{3}{4}e^-$ from the B- to the N-surface leading to a polar catasrophe, what happens here is that the amount of transferred charge depends on $L$, while the quasi-Fermi level offset $\phi$ does not. In the simplest approximation, where the charge transfer leads to degeneracy of the VBM and CBM on opposing sides of the material, one would have $\sigma(L)=\frac{\epsilon_{0}\phi}{L}$. Figure 4(a) shows $\sigma(L)$ as a function of $1/n$ (where $n=L/L_{0}$ with $L_{0}$ being the bulk period along $z$) together with a linear fit. Not only does the linear fit work rather well, but also the fitted $\phi$=2.7 eV is in reasonable agreement with $\phi$=2.4 eV obtained in Fig. 3. We should note that $\sigma(L)$ here includes both the free charge of the surface DBs and the bound charge due to spontaneous polarization. To obtain the amount of free charge transferred across the slab, one would have to subtract the bound charge from $\sigma(L)$ to yield 
\begin{equation}
\sigma_{\text {free }}(L)=\sigma(L)-\sigma_{SP}
\end{equation}
where $\sigma_{SP}$ (calculated before using passivated slabs) has also been placed in Fig. 4(a). Therefore, one can read off the value of $\sigma_{\text {free}}(L)$ as the difference between $\sigma(L)$ and $\sigma_{SP}$. 

As an application to nanomaterials, let us consider the piezoelectric effect, wherein strain alters spontaneous polarization and creates an imbalance between surface free charges to result in a potential difference across a slab. By straining the slab, we can directly calculate the strain effect on the surface net and free charges, as shown for the case of wurtzite $AlN$ in Figs. 4(b) and (c), respectively. One striking feature is that for few atomic layers, the free charge on the surface vanishes and hence becomes independent of the relatively small strain applied here. This is because the surface has been passivated and, as a result, the entire system is insulating. While, as can be seen from the net charge, the strain alters the spontaneous polarization, it does not lead to any charge transfer between the two surfaces. As the system length is increased above approximately 10 atomic layers, however, the spontaneous polarization leads to pinning of the conduction and valence bands of the opposing surfaces, giving rise to an overall system metallicity. To maintain such a pinning, any change in the spontaneous polarization necessitates a change in the free surface charge $\Delta\sigma$ due to strain, from which we calculate piezoelectric constant $d_{33}= \frac{\epsilon_0\Delta c}{c\Delta \sigma}$, where $\frac{\Delta c}{c}$ is the relative change of the lattice constant in the normal $c$ direction.

Figure 4(d) shows the results for $d_{33}$ for AlN. We see that for large $L$, there is relatively weak size dependence, with the calculated $d_{33}$ ranging between 5.05 (as $L\rightarrow\infty$) and 6.26 $pm/V$. These values are in good agreement with experimental values between 5.10 and 6.72 $pm/V$ \cite{30,31,32,33,34,35}. However, change in $d_{33}$ can be quite pronounced in regions where semiconductor/metal transition takes place, as can be seen in Fig. 4(d). This is because for small $L$, the system becomes insulating, and the piezoelectric effect is quenched, leading to $d_{33}\rightarrow\infty$. While such a quenching is generally true as the system becomes insulating, for highly polar semiconductors (or systems with a substantial number of un-passivated DBs) the system may maintain its metallicity, even down to the limit of several atomic layers. For instance, the perovskite $BaTiO_3$, has a spontaneous polarization of 0.26 $c/m^2$, which is nearly 3 times that of $AlN$. As a result, even the self-passivated unit cell, containing 2 atomic layers, is metallic and our calculation yields a non-vanishing $d_{33}$ of 6.06-28.10 $pm/V$. In general, we can only expect a quenching of surface free charge (and the piezoelectric effect) if the potential difference across the sample due to spontaneous polarization is smaller than the band gap, i.e. ($\sigma_{SP}L)/\epsilon_0<E_g$, as schematically shown in Fig. 1(a). 

In summary, we have developed a real-space approach for polarization via calculating the spatial distribution of polarization charge. The method is not only applicable to bulk materials, showing good agreement with Berry-phase calculations, but it can also be applied to systems with DBs and free surface charge. As such, it can greatly contribute to uncovering the rich polarization physics at the nanoscale. For example, we find that in unpassivated nanostructures charge transfer between surface DBs dominates the total polarization of typical wurtzite semiconductors and only in the bulk limit (with a net zero electric field), does the free-charge polarization approaches the spontaneous polarization. As an application, we consider nanoscale piezoelectric effect: not only the calculated piezoelectric coefficients agree with experiment, but we show that for passivated or nearly passivated nano systems below some critical length scale, the surface free charge also vanishes, leading to a complete quenching of the piezoelectric effect.

\begin{acknowledgments}
This material is based upon work supported by the U.S. Department of Energy, Office of Science, Office of Basic Energy Sciences under Award Number DE-SC-0002623. The supercomputer time sponsored by the National Energy Research Scientific Center (NERSC) under DOE Contract No. DE-AC02-05CH11231 and the Center for Computational Innovations (CCI) at RPI are also acknowledged. K. Y. acknowledges the support by China Scholarship Council (CSC). We also gratefully acknowledge discussion with Wei-Qing Huang and Shuangchun Wen.
\end{acknowledgments}

\end{document}


\begin{titlepage}
\title{Supplemenal Material for Polarization at the Nanoscale}
\author{Ke Yang}
\affiliation{Department of Physics, Applied Physics and Astronomy, Rensselaer Polytechnic Institute, Troy, New York 12180 (USA)}
\author{Zeyu Jiang}
\affiliation{Department of Physics, Applied Physics and Astronomy, Rensselaer Polytechnic Institute, Troy, New York 12180 (USA)}
\author{Duk-Hyun Choe}
\affiliation{Samsung Advanced Institute of Technology (SAIT), Suwon, Korea}
\author{Damien West}
\email{damienwest@gmail.com}
\affiliation{Department of Physics, Applied Physics and Astronomy, Rensselaer Polytechnic Institute, Troy, New York 12180 (USA)}
\author{Shengbai Zhang}
\affiliation{Department of Physics, Applied Physics and Astronomy, Rensselaer Polytechnic Institute, Troy, New York 12180 (USA)}
\maketitle
\vspace{2mm}
\end{titlepage}
\section{surface passivation}
In the wurtzite structure, each atom has 4 nearest neighbor (nn) dissimilar atoms and hence 4 bonds, each hosting 2 electrons. Using boron nitride (BN) as an example, by the ECM, each B atom (with three valence electrons) contributes 3/4 = 0.75 electrons to the bond, whereas each N atom (with five valence electrons) contributes 5/4 = 1.25 electrons to the bond. On the (0001) surfaces, one of the four bonds is cut, leaving one dangling bond (DB) on each surface atom. Here, each surface B atom would like to give away its 0.75 electrons, leaving it with an empty DB, while each surface N atom would like to take 0.75 electrons to result in a doubly-occupied DB (i.e., a lone-pair). One can achieve this by having a 2 $\times$ 2 surface reconstruction with one cation vacancy per 2 $\times$ 2 for the (0001) surface, or one anion vacancy per $2 \times 2$ for the $(000\overline{1})$ surface, as shown in Fig. S1 (b), in which case charge transfer happens locally at each individual surface. Alternatively, one can add a pseudo hydrogen (pseudo-H) atom with 1.25 electrons to each surface B atom, or a pseudo-H atom with 0.75 electrons to each surface N atom, to saturate their DBs, as shown in Fig. S1(a). Passivation is a local  perturbation, as adding a charge-neutral object to the surface, such as the pseudo-H atom, does not lead to a long-distance charge transfer across the slab. Hence, the resulting polarization, and the integrated polarization charge, should not depend on the choice of the passivation. Indeed, both passivation schemes, as well as a mixing of the two, yield identical results in our calculation. 

\begin{figure}[tbp]
\includegraphics[width=0.9\columnwidth]{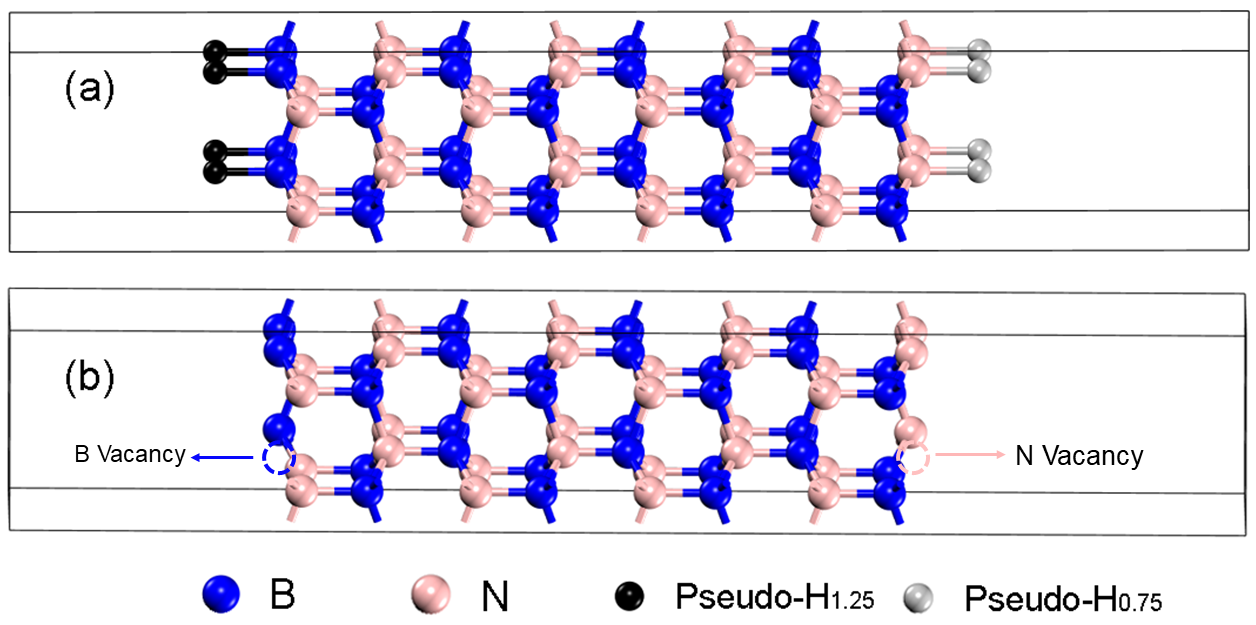}
\caption{\label{fig:S1} Comparison of two passivation schemes for boron nitride (BN). (a) Pseudo hydrogen (associated w/ nuclear and electronic charge of 5/4 or 3/4) is used to satisfy electron counting and saturate surface dangling bonds, leaving each surface semiconducting. (b) A more physical approach to surface passivation in which 1/4 of the surface atoms (either B or N) are replaced with vacancies, leading to semiconducting surfaces which self-compensate, i.e. containing equal numbers of empty B dangling bonds and filled N lone pairs. }
\end{figure}